# Generative AI in EU Law: Liability, Privacy, Intellectual Property, and Cybersecurity

*Working Paper (this version: 15 March 2024)*

Claudio Novelli[1], Federico Casolari[1], Philipp Hacker[2], Giorgio Spedicato[1], Luciano Floridi[1, 3]

[1] Department of Legal Studies, University of Bologna, Via Zamboni, 27/29, 40126, Bologna, IT
[2] European New School of Digital Studies, European University Viadrina, Große Scharrnstraße 59, 15230 Frankfurt (Oder), Germany
[3] Digital Ethics Center, Yale University, 85 Trumbull Street, New Haven, CT 06511, US

*Email of the correspondence author: claudio.novelli@unibo.it[1]

**Abstract:**

The advent of Generative AI, particularly through Large Language Models (LLMs) like ChatGPT and its successors, marks a paradigm shift in the AI landscape. Advanced LLMs exhibit multimodality, handling diverse data formats, thereby broadening their application scope. However, the complexity and emergent autonomy of these models introduce challenges in predictability and legal compliance. This paper analyses the legal and regulatory implications of Generative AI and LLMs in the European Union context, focusing on liability, privacy, intellectual property, and cybersecurity. It examines the adequacy of the existing and proposed EU legislation, including the Artificial Intelligence Act (AIA), in addressing the challenges posed by Generative AI in general and LLMs in particular. The paper identifies potential gaps and shortcomings in the EU legislative framework and proposes recommendations to ensure the safe and compliant deployment of generative models.

**Acknowledgements**: CN's contributions were supported by funding provided by Intesa Sanpaolo to the University of Bologna.

## 1. Overview

Since the release of ChatGPT at the end of 2022, Generative AI in general, and Large Language Models (LLMs) in particular, have taken the world by storm. On a technical level, they can be distinguished from more traditional AI models in various ways.[1] They are trained on vast amounts of text and generate language as output, as opposed to scores or labels in traditional regression or classification (Foster 2022, 4-7; Hacker,

---

[1] While Generative AI encompasses a wider range of systems than LLMs, their overlapping legal concerns necessitate considering them together. However, we will maintain a focus on LLMs.

Engel, and Mauer 2023). Often, Generative AI models are marked by their wider scope and greater autonomy in extracting patterns within large datasets. In particular, LLMs' capability for smooth general scalability enables them to generate content by processing a varying range of input from several domains. Many LLMs are multimodal (also called Large Multimodal Models, LMMs), meaning they can process and produce various types of data formats simultaneously: e.g., GPT-4 can handle text, image, and audio inputs concurrently for generating text, images, or even videos (e.g., Dall-E and Sora integrations). However, while advanced LLMs generally perform well across a broad spectrum of tasks, this comes with unpredictable outputs raising concerns over the lawfulness and accuracy of LLM-generated texts (Ganguli et al. 2022).

As powerful LLMs like GPT-4 and Gemini, image and video generators rise, their very momentum throws into stark relief the question of the adequacy of existing and forthcoming EU legislation. In this article, we discuss some key legal and regulatory concerns brought up by Generative AI and LLMs regarding liability, privacy, intellectual property, and cybersecurity. The EU's response to these concerns should be contextualised within the guidelines of the Artificial Intelligence Act (AIA), which comprehensively addresses the design, development, and deployment of AI models, including Generative AI within its scope. Where we identify gaps or flaws in the EU legislation, we offer some recommendations to ensure that Generative AI models evolve lawfully.

## 2. Liability and AI Act

33% of firms view "liability for damage" as the top external obstacle to AI adoption, especially for LLMs, only rivalled by the "need for new laws", expressed by 29% of companies.[2] A new, efficient liability regime may address these concerns by securing compensation to victims and minimizing the cost of preventive measures. In this context, two recent EU regulatory proposals on AI liability may affect LLMs (Dheu et al. 2022): one updating the existing Product Liability Directive (PLD) for defective products,[3] the other introducing procedures for fault-based liability for AI-related damages through the Artificial Intelligence Liability Directive (AILD).[4] While an interinstitutional agreement has been reached on the text of the new PLD,[5] the AILD is currently parked in the legislative process, but may be taken up again once the AI Act has entered into force.

The two proposals offer benefits for regulating AI liability, including Generative AI and LLMs. First, the scope of the PLD has been extended to include all AI systems and AI-enabled goods, except for open-source software, to avoid burdening research and innovation (Rec. 13 PLD; but see Rec. 13a PLD: covered if integrated into commercial product[6]). This is advantageous as the PLD is the only harmonized European liability law, with a strict liability regime applicable in specific instances. Second, the PLD acknowledges that an AI system can become defective based on knowledge acquired/learned post-deployment, thereby extending liability to such occurrences (Article 6(c) PLD). Third, the AILD covers claims against non-professional users of AI systems and recognizes violations of fundamental rights among eligible damages. Finally, and perhaps most importantly, both proposals acknowledge AI's opacity and the information imbalance between developers and users or consumers. Thus, they introduce disclosure mechanisms and rebuttable presumptions, shifting the burden of proof to providers or deployers (AILD Articles 3 and 4; PLD Articles 8 and 9). For instance, under Article 8 PLD and Article 3 AILD, claimants only need to provide plausible evidence of potential harm, while defendants must disclose all relevant information to avoid liability, with non-compliance to this disclosure leading to a (rebuttable) presumption that the defendant has breached its duty of care.

However, both AILD and PLD reveal three major weaknesses (see below) when used in the context of Generative AI, largely stemming from their dependence on the AI Act (AIA), which appears ill-suited to govern LLMs effectively. Although the text of the AIA is now stable, it is important to consider improvements in the next legislative phases, such as the comitology procedure enabling implementing acts, before the AIA is enforced, which is expected to happen no earlier than 2026. For Generative AI and LLMs – labelled as General Purpose AI (GPAI) models – obligations will apply sooner, specifically 12 months after the AIA's entry into force. For existing GPAIs on the market when the AI Act rules are applied, this transition period is extended to 24 months (Art. 83(3) AIA).

1) *Scope.* The disclosure mechanism and rebuttable presumption of a causal link in the AILD only apply to high-risk AI systems under the AIA. Hence, the primary issue here is to establish whether, and under what conditions, Generative AI (and LLMs) might fall under the scope of the AILD and its liability mechanism.

During the drafting of the AIA, GPAI models were first classified as high-risk by default. Subsequently, the risk assessment shifted to consider their downstream application (e.g., if used in a high-risk context such as a judicial settings). Finally, the consolidated version has provided a distinct classification. They carry a set of distinct, overarching obligations (Articles 53 ff., AIA). This framework introduces a tiered risk classification that diverges from the traditional high, medium, or low-risk categories: (1) providers of *standard* GPAI must always ensure detailed technical and informational documentation, also to enable downstream users to comprehend their capabilities and limitations, intellectual property law adherence (e.g., copyright Directive), and transparency about training data (Article 53, AIA); (2) providers of *openly licensed* GPAI models, i.e., with publicly accessible parameters and architecture, need only meet technical documentation requirements (Article 53, point 2); (3) providers of GPAI models posing *systemic risks* must fulfil standard obligations and additionally conduct model evaluations, including adversarial testing (red teaming), assess and mitigate risks, document and report incidents to the AI Office, and maintain adequate cybersecurity protection (Article 55(c), AIA). This also applies to open-source models. A GPAI model is considered to pose systemic risks if the Commission, either on its initiative or based on recommendations from the scientific panel, recognizes it as having high-impact capabilities. This recognition must be based on specific technical metrics and is automatically presumed if the model's training involves more than $10^{25}$ floating-point operations (FLOPs). The rationale behind using FLOPs as a benchmark is the belief that higher computational resources indicate more sophisticated models, which may have broader impacts on society.[7]

---

[6] See the corresponding policy suggestion and argument made in (Hacker 2023a, at footnote 107).

The Commission advises providers of GPAI models with systemic risks to create a code of conduct with expert help, demonstrating compliance with the AI Act. This is especially important for outlining how to assess and manage risks for GPAI models with systemic risks. As a result, GPAI with systemic risks are likely to be subjected to the disclosure mechanism and rebuttable presumption in the AILD.

While these revisions to the AIA represent a positive step toward more effective risk assessment, concerns remain. So, for instance, the three-tier classification system to GPAIs – standard, open licensed, and systemically risky – may fail to account for the peculiarities of downstream applications, potentially leading to over-inclusive or under-inclusive risk categories (Novelli et al. 2024; 2023). The same definition of systemically risky GPAI models, primarily based on the computational resources used for training (FLOPs), may not capture their multidimensional nature: they depend on various factors such as the context of application, model architecture, and the quality of training, rather than just the quantity of computational resources used. FLOPs offer only a partial perspective on dangerousness and do not account for how different, non-computational risk factors might interact and potentially lead to cascading failures, including interactions among various LLMs. Finally, the very threshold of $10^{25}$ FLOPs as a risk parameter is questionable (The Future Society 2023) (Moës and Ryan 2023). LLMs with $10^{24}$ or $10^{23}$ FLOPs can be equally risky (e.g., GPT-3; Bard). This is further compounded by the trend towards downsizing LLMs while maintaining high performance and associated risks, such as in the case of Mistral's Mixtral 8x7B model (Hacker 2023b). Again, while this is an ancillary issue as the AI Office will have the power to adjust this parameter, relying solely on FLOPs as a risk indicator remains inadequate.

The AILD, proposed in September 2022, predates the drafting process of the final text of the AIA, which has undergone significant changes, particularly with the rise of LLMs in 2023. Therefore, it is necessary to update the AILD to align with the new technologies, risk categories, and obligations introduced in the AIA. A question arises regarding which type of Generative AI models the disclosure and rebuttable presumption mechanism should apply to. Given that all providers of GPAI models, including those with open licenses, will be subject to rigorous transparency and recordkeeping obligations, it seems reasonable to extend the disclosure mechanism and rebuttable presumption of causal link to all of them. This is because they are assumed to have the necessary information in case of incidents, and their failure to provide it can be used as a presumption of violation of the standards set by the same

---

[7] However, the Commission must adjust the threshold as technology advances, like better algorithms or more efficient hardware, to stay current with the latest developments in general purpose AI models.

AIA. However, the AILD's liability rules may prove overly stringent for some GPAI models, suggesting the need for exemptions. To facilitate these exemptions, additional criteria for classifying GPAI models are necessary.[8] In a similar vein, the AI Act introduces criteria that prevent AI systems operating in Annex III from being automatically deemed high-risk; they must instead present a *significant risk* to people or the environment. Likewise, Article 7 of the AIA empowers the Commission to adjust the high-risk designation by adding or removing specific applications or categories. A similar approach for GPAI could exempt certain Generative AI models from AILD's strict requirements. This could involve tailoring the three-tier classification to real-word Generative AI risk scenarios (Novelli et al. 2024; 2023), based not only to computation potency, but on their specific deployment contexts, considering the potential harms to assets and individuals (Bender et al. 2021).[9] For example, in the employment sector — deemed high-risk by the AIA — the risk levels can significantly differ between using LLMs just for resume screening optimization or for automated virtual interviews, where biases could be more common and human oversight less effective. Alternatively, exemptions for GPAI models could be established by aligning the three-tier system with the broad application areas designated for AI systems (e.g., Annex III). This way, models used in lower-risk areas, such as video games, could be exempted from the AILD's more stringent liability rules.

2) *Defectiveness and fault*. The two directive proposals assume that liability may arise from two different sources–defectiveness (PLD) and fault (AILD)–that are both evaluated by compliance with the requirements of the AIA. Both presume fault/a defect in case of non-compliance with the (high-risk systems) requirements of the AIA (Article 9(2)(b) PLD; Article 4(2) AILD), requirements which could also be introduced at a later stage by sectoral EU legal instruments.[10] However, these requirements may not be easily met during the development of Generative AI, particularly LLMs: e.g., their lack of a single or specific purpose before adaptation (Bommasani et al. 2022) could hamper the predictions of their concrete impact on the health, safety, and fundamental rights of persons in the Union which are required by the AIA risk management system and transparency obligations (Articles 9 and 13, AIA). Moreover, as just mentioned, further requirements are likely to be introduced in the EU regulatory framework concerning GPAI models.

To enhance the effectiveness and reliability of Generative AI models, a necessary recommendation is to combine the conventional AI fault and defectiveness criteria with new methods specifically designed to align with their technical nuances. This may imply that the compliance requirements for evaluating faults and defectiveness should prioritize techniques for steering the randomness of their non-deterministic outputs over their intended purposes. Indeed, their capability for smooth general scalability enables them to generate content by processing diverse inputs from arbitrary domains with minimal training (Ganguli et al. 2022). To this scope, several techniques might be incentivised by the regulator, also concurrently: e.g., temperature scaling, top-k sampling, prompt engineering, and adversarial training (Hu et al. 2018). Methods for tempering the randomness may also include the so-called regularization techniques, like the dropout, which involves temporarily disabling a random selection of neurons during each training step of Generative AI models, fostering the development of more robust and generalized features (Lee, Cho, and Kang 2020). Consequently, it prevents the model from overfitting, ensuring more coherent and less random outputs.

---

[8] These can introduced both in the Commission's delegated acts and throughout the standardization process.
[9] The PLD, which is not tied to the risk categories of the AIA in terms of applicability, cannot do all the work because its provisions apply only to professionals – economic operators – and not to non-professional users like the AILD.
[10] The dependence on the AIA is less of an issue for the PLD as it has greater harmonization and extensive case law. However, identifying the appropriate safety requirements (Articles 6 and 7) to assess the defectiveness of Generative AI and LLMs remains a challenge.

Furthermore, compliance requirements for Generative AI and LLMs should also prioritize monitoring measures. These measures would serve to verify that the models operate as planned and to pinpoint and amend any divergences or unfavourable results. For example, calculating the uncertainty of outputs could be instrumental in recognizing situations where the model might be producing highly random results (Xiao et al. 2022). Such information is vital for end-users to have before utilizing, for instance, an LLM, representing a metric for evaluating the fault of the designers and deployers (or the defectiveness of the same).

3) *Disclosure of evidence*. Both proposals state that the defendant — in our analysis, the deployers and designers of a Generative AI model — must provide evidence that is both relevant and proportionate to the claimant's presented facts and evidence. Shortcomings here concern the content of such disclosure. First, the PLD and the AIA are misaligned as the former requires evidence disclosure for all AI systems, whereas the AIA proposal mandates record-keeping obligations only for high-risk systems (Article 12, AIA) (Hacker 2023a). Regarding Generative AI, there is no blanket requirement for GPAI providers to continuously and automatically record events ('logging') throughout the model's lifecycle. The obligation to document and report significant incidents to the AI Office and national authorities is limited to models classified as systemically risky (Article 55c, AIA). Providers of standard GPAI are required to maintain technical documentation related to training, testing, model evaluation outcomes, and proof of copyright law compliance, yet there is no directive for ongoing performance recording.

Second, both the PLD and the AILD do not indicate what type of information must be disclosed. While this issue can be attributed to their status as proposals, it is this gap that should be promptly addressed. Failing to establish clear guidelines on the necessary disclosures might leave the claimants practically unprotected.

Regarding the first issue, the requirement to disclose evidence should not be confined to high-risk systems alone. The PLD could potentially adopt the AILD's approach, which broadens the disclosure requirement to include opaque systems that are not classified as high-risk while exempting those high-risk systems that already have ample documentation under the AIA (Article 4(4) and (5), AILD). This strategy could broaden the scope to include standard GPAI models, not just those systemically risky. This adjustment is reasonable, particularly since standard GPAI models typically process less data than their systemically risky counterparts and already face stringent transparency obligations that should facilitate the implementation of record-keeping practices. While the content of disclosure might vary based on the system's risk level, maintaining the obligation to disclose is important.

This leads us to the second point of discussion: the content of disclosure. It should include a report of the damaging incident, noting the exact time and a brief description of its nature. It might include interaction logs and timestamps between users and the GPAI model, demonstrating adherence to relevant standards, possibly verified through third-party audit reports (Falco et al. 2021). Moreover, the disclosure should also mirror the sociotechnical structure of AI liability (Novelli, Taddeo, and Floridi 2023; Theodorou and Dignum 2020) and demonstrate that training data are representative and well-documented, e.g., in terms of the motivation behind data selection and transparency about the objectives of data collection (Bender et al. 2021; Jo and Gebru 2020). Also, producers might be obligated to use only documentable datasets of an appropriate size for the capabilities of the organization. For instance, LLMs operating on restricted datasets–thanks to their few/zero-shot learning skills (Brown et al. 2020)–may instead need to disclose the auxiliary information used for associating observed and non-observed classes of objects.

To conclude, the process of evidential disclosure presupposes that individuals are informed when they are engaging with these models, and consequently, whether they have been adversely affected in specific ways. However, even though the stipulations outlined in the AIA mandate the notification of users during interactions with GPAI models, the methodology for user notification remains ambiguous (Ziosi et al. 2023). This is a key point as the efficacy of the disclosure mechanisms hinges on this prerequisite, wherein to lodge claims, users must possess a reasonable basis to suspect harm and furnish substantial details and corroborating evidence to substantiate a potential damages claim. Since the acquisition of this knowledge can present challenges, it is recommended to encourage Generative AI producers to actively notify occurrences of potential harm. This strategy would not only bolster the claimant's ability to access crucial evidence but would also cultivate a more transparent environment within the operational sphere of Generative AI models. Such incentives might include initiatives like forming alliances with credible third-party organizations, including auditing agencies, to facilitate a thorough disclosure of information (and evidence) concerning adverse effects.

## 3. Privacy and Data Protection

Privacy and data protection pose critical legal hurdles to the development and deployment of Generative AI, as exemplified by the 2023 Italian data authority's (Garante della Privacy) temporary ban on ChatGPT (Hacker, Engel, and Mauer 2023, Technical Report) and the following notice in January 2024 by the same authority to OpenAI that its ChatGPT chatbot allegedly violates the EU General Data Protection Regulation. On an abstract level, a Generative AI models preserves privacy if it was trained in a privacy-sensitive way, processes prompts containing personal data diligently, and discloses information relating to identifiable persons in appropriate contexts and to authorized individuals only. Privacy and data protection are not binary variables and, therefore, what is the right context or the right recipients of the information is a matter of debate. In the context of LLMs, these debates are further complicated due to the diverse purposes, applications, and environments they operate in.[11]

Generative AI models are exposed to privacy and data protection violations due to pervasive training on (partially) personal data, the memorization of training data, inversion attacks (Nicholas Carlini et al. 2021), interactions with users, and the output the AI produces. Memorization of data may occur either through overfitting of abundant parameters to small datasets, which reduces the capacity to generalize to new data, or through the optimizing generalization of long-tailed data distributions (Feldman 2021). When the memorized training data contains personal information, LLMs may leak data and disclose it directly. When training data is not memorized, personal information can still be inferred or reconstructed by malicious actors using model inversion attacks, which reverse-engineer the input data to reveal private information (Fredrikson, Jha, and Ristenpart 2015). Against this, the existing privacy-preserving strategies, such as data sanitization and differential privacy, provide limited privacy protection when applied to LLMs (Brown et al. 2022). This raises the question of whether, and how, personal data may be processed to train LLMs–a particularly thorny question concerning sensitive data. Moreover, users may enter private information through prompts, which may resurface in other instances. Some users, in addition, will be minors, for whom specific data protection rules apply.

These considerations lead to seven main problems at the intersection of data protection and Generative AI:[12] the appropriate legal basis for AI training; the appropriate legal basis for processing prompts; information requirements; model inversion, data leakage, and the right to erasure; automated decision-making; protection of minors; and purpose limitation and data minimization. We analyse them first and then offer some thoughts on potential ways forward.

1) *Legal basis for AI training on personal data.* First and foremost, every processing operation of personal data–be it storage, transfer, copying, or else–needs a legal basis under Article 6 GDPR. For companies without an establishment in the EU, the GDPR still applies if their services are offered in the EU, for example, which is the case for many major LLM products. The GDPR also covers processing before the actual release of the model, i.e., for training purposes (Oostveen 2016). LLMs are typically trained on broad data at scale, with data sources ranging from proprietary information to everything available on the Internet–including personal data, i.e., data that can be related to an identifiable individual (Bommasani et al. 2021). Using this type of data for AI training purposes, hence, is illegal under the GDPR unless a specific legal basis applies. The same holds for any fine-tuning operations after initial pre-training.

*1.a) Consent and the balancing test*

---

[11] For this discussion, we will concentrate on strategies to prevent LLMs from compromising user privacy and personal data, bypassing what makes a context or a recipient. However, an analysis of these issues is done by (Brown et al. 2022).

[12] This list is not exhaustive. For practitioners, particularly, the records of processing activities (Article 30 GDPR) and the data protection impact assessment (Article 35 GDPR) are very relevant as well. See, e.g., the data protection checklist for AI issued by the Bavarian Data Protection Authority, https://www.lda.bayern.de/media/ki_checkliste.pdf.

The most prominent legal basis in the GDPR is consent (Article 6(1)(a), GDPR). However, for large data sets including personal information from a vast group of people unknown to the developers beforehand, eliciting valid consent from each individual is generally not an option due to prohibitive transaction costs (Mourby, Ó Cathaoir, and Collin 2021). Furthermore, using LLMs with web-scraped datasets and unpredictable applications is difficult to square with informed and specific consent (Bommasani et al. 2022). At the same time, requiring data subjects to be informed about the usage of their personal data may slow down the development of LLMs (Goldstein et al. 2023). Hence, for legal and economic reasons, AI training can typically be based only on the balancing test of Article 6(1)(f) GDPR (Zuiderveen Borgesius et al. 2018; Zarsky 2017), according to which the legitimate interests of the controller (i.e., the developing entity) justify processing unless they are overridden by the rights and freedoms of the data subjects (i.e., the persons whose data are used).[13]

Whether the balancing test provides a legal basis is, unfortunately, a matter of case-by-case analysis (Gil Gonzalez and de Hert 2019; Peloquin et al. 2020; Donnelly and McDonagh 2019). Generally, particularly socially beneficial applications will speak in favour of developers; similarly, the data subject is unlikely to prevail if the use of the data for AI training purposes could reasonably be expected by data subjects, Recital 47. That latter criterion, however, will rarely be fulfilled. In addition, privacy-enhancing strategies, such as pseudonymization, transparency or encryption, will count toward the legality of AI training under the balancing test. By contrast, the nature and scope of processing, the type of data (sensitive or not), the degree of transparency towards and control for data subjects, and other factors may tip the balance in the other direction (Hacker, Engel, and Mauer 2023, Technical Report, 2).

For narrowly tailored AI models based on supervised learning strategies, one may argue that AI training is not particularly harmful as it does not, generally, reveal any new information about the data subjects themselves (Hacker 2021; Zarsky 2017; Bonatti and Kirrane 2019). This argument is particularly strong if the model is not passed along to other entities and state-of-the-art IT security makes data breaches less likely.

However, this position is difficult to maintain concerning Generative AI (Hacker, Engel, and Mauer 2023, Technical Report, 2): these models are generally used by millions of different actors, and models have been shown to reveal personal data through data leakage as well as model inversion (Nicholas Carlini et al. 2021; Bederman 2010; Lehman et al. 2021; Nicolas Carlini et al. 2023). This poses an even greater challenge in fine-tuning scenarios (Borkar 2023).

*1.b) Sensitive Data*
To make matters even more complex, a much larger number of personal data pieces than expected may be particularly protected as sensitive data pursuant to Article 9 GDPR, under a new ruling of the CJEU. In Meta v. Bundeskartellamt, the Court decided that information need not directly refer to protected categories–such as ethnic or racial origin, religion, age, or health–to fall under Article 9. Rather, it suffices "that data processing allows information falling within one of those categories to be revealed".[14] That case was decided concerning Meta, the parent company of Facebook, based on its vast collection of data tracking users and linking that data with the user's Facebook account.

---

[13] Another possibility is the purpose change test (Article 6(4) GDPR), not explored further here for space constraints. Note that Article 9 GDPR, in our view, applies in addition.

Arguably, however, as is generally the case in technology-neutral data protection law, the exact method of tracking or identification is irrelevant; the Court held that it does not matter, for example, whether the profiled person is a Facebook user or not.[15] Rather, from the perspective of data protection law, what is decisive is the controller's ability to infer sensitive traits based on the available data–irrespective of whether the operator intends to make that inference. This broader understanding casts a wide net for the applicability of Article 9 GDPR, as machine learning techniques increasingly allow for the deduction of protected categories from otherwise innocuous data points (Bi et al., 2013; Chaturvedi & Chaturvedi, 2024).

Hence, in many cases concerning big data formats, the hypothetical possibility to infer sensitive data potentially brings the processing, for example, for AI training purposes, under the ambit of Article 9. Developers then need to avail themselves of the specific exception in Article 9(2) GDPR. Outside of explicit consent, such an exception will, however, often not be available: Article 9(2) does not contain a general balancing test, in contrast to Article 6(1) GDPR (and the secondary use clause in Article 6(4)). The research exemption in Article 9(2)(j) GDPR, for example, is limited to building models for research purposes, but cannot be used to exploit them commercially (cf. Recitals 159 and 162).

Overall, this discussion points to the urgent need to design a novel exemption to Article 9, accompanied by strong safeguards, similar to the ones contemplated in Article 10, point 5 of the AIA, to balance the societal interest in socially beneficial AI training and development with the protection of individual rights and freedoms, particularly in crucial areas such as medicine, education, or employment. While the TDM exception provides for a specific framework for the use of copyrighted material for AI training purposes (see below, 4), such rules are, unfortunately, entirely lacking under the GDPR.

2) *Legal basis for prompts containing personal data.*
The situation is different for prompts containing personal data entered into a trained model. Here, we have to fundamentally distinguish two situations. First, users may include personal information about themselves in prompts, for example, when they ask an LLM to draft an email concerning a specific event, appointment, or task. This may occur intentionally or inadvertently. In both cases, consent may indeed work as a legal basis as users have to individually register for the LLM product. During that procedure, controllers may request consent (respecting the conditions for valid consent under Articles 4(11) and 7 GDPR, of course).

The second scenario concerns prompts containing personal information about third parties, i.e., not the person entering the prompt. This situation is more common among users who might not be fully aware of privacy and data protection laws. They might inadvertently include the personal details of others if the task at hand involves these third parties, and they expect the language model to provide personalized

---

responses. Users cannot, however, validly consent for another person (unless they have been explicitly mandated by that person to do just that, which is unlikely).

A similar problem resurfaces as in the AI training or fine-tuning scenario, with the additional twist that the information is provided, and processing initiated, by the user, not the developers. While the user may be regarded as the sole controller, or joint controller together with the company operating the LLM (Article 4(7) GDPR), for the initial storage and transfer of the prompt (i.e., writing and sending the prompt), any further memorization or data leakage is under the sole control of the entity operating the LLM. Hence, under the *Fashion ID* judgment of the CJEU,[16] that operational entity will likely be considered the sole controller, and thus the responsible party (Art. 5(2) GDPR), for any storage, transfer, leakage, or other processing of the third-party-related personal data included in the prompt that occurs after the initial prompting by the user. Again, as in the training scenario, both the third-party-related prompt itself and any additional leakage or storage are difficult to justify under Article 6(1)(f) and, if applicable, Article 9 GDPR.

### 3) *Information requirements.*

The next major roadblocks for GDPR-compliant Generative AI models are Articles 12-15 GDPR, which detail the obligations regarding the information that must be provided to data subjects. These articles pose a unique challenge for Generative AI due to the nature and scope of data they process (Hacker, Engel, and Mauer 2023, Technical Report, 2-3).

When considering data harvested from the internet for training purposes, the applicability of Article 14 of the GDPR is crucial. This article addresses the need for transparency in instances where personal data is not directly collected from the individuals concerned. However, the feasibility of individually informing those whose data form part of the training set is often impractical due to the extensive effort required, potentially exempting it under Article 14(5)(b) of the GDPR. Factors such as the volume of data subjects, the data's age, and implemented safeguards are significant in this assessment, as noted in Recital 62 of the GDPR. The Article 29 Working Party particularly notes the impracticality when data is aggregated from numerous individuals, especially when contact details are unavailable (Article 29 Data Protection Working Party 2018, para. 63, example).

Conversely, the processing of personal data submitted by users on themselves in a chat interface (prompts) is not subject to such exemptions. Article 13 of the GDPR explicitly requires that data subjects be informed of several key aspects, including processing purposes, the legal basis for processing, and any legitimate interests pursued. Current practices may not have fully addressed these requirements, marking a significant gap in GDPR compliance.

Importantly, the balance between the practical challenges of compliance and the rights of data subjects is delicate. While the concept of disproportionate effort under Article 14(5) GDPR presents a potential exemption, it remains a contentious point, particularly for training data scraping and processing for commercial purposes. In this regard, the data controller, as defined in Article 4(7) of the GDPR, should meticulously document the considerations made under this provision. This documentation is a crucial aspect of the accountability principle enshrined in Article 5(2) of the GDPR. Furthermore, in our view, documents regarding the methods of collecting training data should be made publicly accessible, reinforcing the commitment to GDPR principles.

---

[16] CJEU, C-40/17, Fashion ID, ECLI:EU:C:2019:629.

*4) Model inversion, data leakage, and the right to erasure.*
GDPR compliance for Generative AI models gets even trickier with concerns about reconstructing training data from the model (model inversion) and unintentional data leaks, especially in light of the right to be forgotten (or right to erasure) under Article 17. Some scholars even argue that LLMs themselves might be considered personal data due to their vulnerability to these attacks (Veale, Binns, and Edwards 2018). Inversion attacks refer to techniques whereby, through specific techniques, individuals' data used in the training of these models can be extracted or inferred. Similarly, the memorization problem, which causes LLMs to potentially output personal data contained in the training data, may be invoked to qualify LLMs themselves as personal data.

The ramifications of classifying the model as personal data are profound and far-reaching. If an LLM is indeed deemed personal data, a legal basis is needed for even using or downloading the model. Furthermore, such a qualification implies that data subjects could, in theory, invoke their right to erasure under Article 17 of the GDPR with respect to the entire model. This right, also known as the 'right to be forgotten,' allows individuals to request the deletion of their personal data under specific conditions. In the context of LLMs, this could lead to unprecedented demands for the deletion of the model itself, should it be established that the model contains or constitutes personal data of the individuals.

Such a scenario poses significant challenges for the field of AI and machine learning. The practicality of complying with a request for erasure in this context is fraught with technical and legal complexities (Villaronga et al., 2018; Zhang et al., 2023). Deleting a model, particularly one that has been widely distributed or deployed, could be technologically challenging and may have significant implications for the utility and functionality of the corresponding AI system. Furthermore, this approach raises questions about the balance between individual rights and the broader benefits of AI technologies. The deletion of entire models, with a potential subsequent economic need to retrain the entire model, also conflicts with environmental sustainability given the enormous energy and water consumption of (re-)training LLMs (Hacker 2024).

Although LLM producers, such as OpenAI, claim to comply with the right to erasure, it is unclear how they can do so because personal information may be contained in multiple forms in an LLM, which escalates the complexity of identifying and isolating specific data points, particularly when the data is not presented in a structured format (e.g., phone numbers). Additionally, the removal requests initiated by a single data subject may prove to be inadequate, especially in scenarios where identical information has been circulated by multiple users during their engagements with the LLM (Brown et al. 2022). In other words, the deletion of data from a training dataset represents a superficial solution, as it does not necessarily obliterate the potential for data retrieval or the extraction of associated information encapsulated within the mode's parameters. Data incorporated during the training phase can permeate the outputs generated by certain machine learning models, creating a scenario where original training data, or information linked to the purged data, can be inferred or "leaked," thereby undermining the integrity of the deletion process and perpetuating potential privacy violations (De Cristofaro 2020). At a minimum, this points to the need for more robust and comprehensive strategies to address data privacy and "machine unlearning" (Hine et al., 2023; Floridi 2023; Nguyen et al., 2022) within the operational area of LLMs.

*5) Automated decision-making.*
Furthermore, given new CJEU jurisprudence, the use of Generative AI models might be qualified as automated decision-making processes, a topic scrutinized under Article 22 of the GDPR. This article generally prohibits automated individual decision-making, including profiling, which produces legal effects concerning an individual or similarly significantly affects them, unless specific exceptions apply.

In cases where LLMs are used for evaluation, such as in recruitment or credit scoring, the importance of this regulation becomes even more significant. A pertinent illustration is provided by the recent ruling in the SCHUFA case by the CJEU.[17] The Court determined that the automated generation of a probability value regarding an individual's future ability to payment commitments by a credit information agency constitutes 'automated individual decision-making" as defined in Article 22. According to the Court, this presupposes, however, that this probability value significantly influences a third party's decision to enter into, execute, or terminate a contractual relationship with that individual.

Extrapolating from this ruling, the automated evaluation or ranking of individuals by LLMs will constitute automated decision-making if it is of paramount importance for the decision at hand—even if a human signs off on it afterward. The legal implications of this judgment are profound. Exemptions from the general prohibition of such automated decision-making are limited to scenarios where there is a specific law allowing the process, explicit consent, or where the automated processing is necessary for contractual purposes, as per Article 22(2) of the GDPR.

Obtaining valid consent in these contexts is challenging, especially considering the power imbalances often present between entities like employers or credit agencies and individuals seeking jobs or credit (Recital 43 GDPR). Therefore, the legality of using LLMs in such situations may largely depend on whether their use can be justified as necessary for the specific task at hand (Article 22(2)(a) GDPR). arguments based solely on efficiency are unlikely to be sufficient. Instead, those deploying LLMs for such purposes might need to demonstrate tangible benefits to the applicants, such as more reliable, less biased, or more transparent evaluation processes. Absent such a qualification, only specific union or Member State laws, containing sufficient safeguards, may permit such automated decision making (Article 22(2)(b) GDPR).

*6) Protection of minors.*
The deployment of Generative AI models has raised significant concerns regarding age-appropriate content, especially given the potential for generating outputs that may not be suitable for minors. Under Article 8(2) GDPR, the controller must undertake "reasonable efforts to verify […] that [children's] consent is given or authorized by the

---

[17] CJEU, C-634/21, QG vs. SCHUFA, ECLI:EU:C:2023:957, para. 73.

holder of parental responsibility over the child, taking into consideration available technology."

A notable instance of regulatory intervention in this context is the action taken by the Italian Data Protection Authority (Garante per la Protezione dei Dati Personali—GPDP). On March 30, 2023, the GPDP imposed a temporary restriction on OpenAI's processing of data from Italian users, with a particular emphasis on safeguarding minors.[18] This move underscores the increasing scrutiny by data protection authorities on the implications of LLMs in the context of protecting vulnerable groups, especially children (Malgieri 2023).

In response to these concerns, OpenAI, for example, has implemented measures aimed at enhancing the protection of minors. These include the establishment of an age gate and the integration of age verification tools. The effectiveness and robustness of these tools, however, remain an area of keen interest and ongoing evaluation, especially in the rapidly evolving landscape of AI and data protection.

### 7) *Purpose limitation and data minimization.*
Data controllers should collect personal data only as relevant and necessary for a specific purpose (Article 5(b)-(c), GDPR). The AIA reflects this, requiring an assessment of data quantity and suitability (Article 10, point 2(e), AIA). However, limiting Generative AI models' undefined range of purposes, which need extensive data for effective training, might be futile and counterproductive.

One approach to address data calibration for open-ended LLM applications is requiring developers to train models on smaller datasets and leverage few/zero-shot learning skills. As an alternative to imposing restrictions on the dataset, however, it could be more beneficial to strengthen privacy-preserving measures proportionally to dataset size. For example, rather than relying solely on pseudo-anonymization and encryption (Article 10, point 5(b), AIA), LLM providers should implement methods like differential privacy to counter adversarial attacks on large datasets (Shi et al. 2022; Plant, Giuffrida, and Gkatzia 2022).

### 8) *Ways forward.*
To enable Generative AI models to comply with GDPR data protection standards, we have already suggested a tailored regime under Art. 9(2) GDPR above. Another reasonable step would be to adapt the data governance measures outlined for high-risk systems in the AIA. The Europea Parliament had made proposal for an Article 28(b), which would have delineated the following obligations for GPAI providers: "process and incorporate only datasets that are subject to appropriate data governance measures […] in particular measures to examine the suitability of the data sources and possible biases and appropriate mitigation". However, this proposal has not made it into the final version of the AI Act; rather, if used in specific high-risk scenarios, GPAIs will fall under the data governance rules of Article 10.

While the revised iteration of the compromise text for Article 10 is extensive, it may also be too generic, necessitating the incorporation of more tailored measures or incentives to aptly address the complexities inherent to GPAI models like LLMs (e.g., under harmonised standards and common specifications, Art. 40 and 41 AIA). These technical standards should be refined by incorporating LLM-specific measures, such as requiring training on publicly available data, wherever possible. A significant portion of these datasets might also take advantage of GDPR's right to be forgotten exceptions for public interest, scientific, and historical research (Article 17(3)(d)). Where these exceptions do not apply, it could be feasible for LLMs to use datasets not contingent upon explicit consent, which are intended for public usage. Hence, the most appropriate way to use these systems could require fine-tuning public data with private information for individual data subjects' local use. This should be allowed to maximize LLMs' potential, as proposed by (Brown et al. 2022).

---

[18] Garante per la Protezione dei Dati Personali, Provvedimento del 30 marzo 2023 [9870832].

Other potential strategies to enhance data privacy are: encouraging the proper implementation of the opt-out right by LLM providers and deployers and exploring the potential of machine unlearning (MU) techniques, as mentioned.

Regarding the first strategy, OpenAI has recently made a potentially significant advancement in this direction by releasing a web crawler, named GPTbot, that comes with an opt-out feature for website owners. This feature enables them to deny access to the crawler, as well as customize or filter accessible content, granting them control over the content that the crawler interacts with.[19] This is useful not only for implementing the opt-out right under the EU TDM copyright exception but also under Article 21 GDPR.

Turning to the second strategy, MU stands as potentially a more efficient method to fully implement the right to erasure (Nguyen et al. 2022), a critical aspect when dealing with LLMs. Unlike conventional methods that merely remove or filter data from a training set — a process that is often inadequate since the removed data continues to linger in the model's parameters — MU focuses on erasing the specific influence of certain data points on the model, without the need for complete retraining. This technique, therefore, could more effectively enhance both individual and group privacy when using LLMs (Hine et al. 2023; Floridi 2023).

## 4. Intellectual Property

Next to data protection concerns, Generative AI presents various legal challenges related to its "creative" outputs. Specifically, contents generated by LLMs result from processing text data such as websites, textbooks, newspapers, scientific articles, and programming codes. Viewed through the lens of intellectual property (IP) law, the use of LLMs raises a variety of theoretical and practical issues[20] that can only be briefly touched upon in this paper, and that the EU legislation seems not yet fully equipped to address. Even the most advanced piece of legislation currently under consideration by the EU institutions – the AIA – does not contain qualified answers to the issues that will be outlined below. The stakes have been raised significantly, however, by several high-profile lawsuits levelled by content creators (e.g., the New York Times;

---

[19] However, skepticism about opting-out tools has raised because, for example, individual users opting-out are not the only holder of their sensitive information (Brown et al. 2022).

[20] For a general discussion of these issues, see (J.-A. Lee, Hilty, and Liu 2021) and the compendium provided by WIPO, *Revised Issues Paper on Intellectual Property Policy and Artificial Intelligence*, 21 May 2020, WIPO/IP/AI/2/GE/20/1 REV.

Getty Images) against Generative AI developers, both in the US[21] and in the EU (de la Durantaye 2023).

Within the context of this article, it is advisable to distinguish between the training of LLMs and the subsequent generation of outputs. Furthermore, concerning the generation of outputs, it is worthwhile to further differentiate–as suggested, among the others, by the European Parliament[22] – between instances in which LLMs serve as mere instruments to enhance human creativity and situations in which LLMs operate with a significantly higher degree of autonomy. On the contrary, the possibility of protecting LLMs themselves through an IP right will not be discussed in this paper.

1) *Training.*
The main copyright issue concerning AI training arises from the possibility that the training datasets may consist of or include text or other materials protected by copyright or related rights (Sartor, Lagioia, and Contissa 2018). Indeed, for text and materials to be lawfully reproduced (or otherwise used within the training process), either the rightholders must give their permission or the law must specifically allow their use in LLM training.

The extensive scale of the datasets used and, consequently, the significant number of rightholders potentially involved render it exceedingly difficult to envision the possibility that those training LLMs could seek (and obtain) an explicit license from all right holders, reproducing the problem of data protection consent. This issue becomes particularly evident when, as often occurs, LLM training is carried out using web scraping techniques, a practice whose legality has been (and continues to be) debated by courts and scholars in Europe (Sammarco 2020; Klawonn 2019), even in terms of potential infringement of the *sui generis* right granted to the maker of a database by Directive 96/9/EC[23]. On the one hand, some content available online, including texts and images, might be subject to permissive licensing conditions–e.g. some Creative Commons licenses–authorizing reproduction and reuse of such content even for commercial purposes. The owner of a website could, on the other hand, include contractual clauses in the Terms and Conditions of the website that prohibit web scraping even when all or some of the website's content is not *per se* protected by intellectual property rights.[24] To mitigate legal risk, LLMs should be suitably capable of autonomously analyzing website Terms and Conditions, thereby discerning between materials whose use has not been expressly reserved by their rightholders and materials that may be freely used (also) for training purposes.

The OpenAI above's GPTbot web crawler which allows website owners to opt-out or filter/customize content access offers a significant technical tool in this context. While it does not eliminate all IP law concerns, it is a proactive measure that could, in the future, set a standard of care that all LLMs' providers might be expected to uphold.[25] Significantly, the GPAI rules of the AIA discussed in the trilogue contained precisely an obligation for providers of such systems to establish a compliance system, via technical and organizational measures, capable of recognizing and respecting rightholders' opt-outs (Hacker 2023c). For the moment, it remains unclear, however, if this provision will be contained in the final version of the AI Act. It would be a step in the right direction, as commercial LLM training without such a compliance system typically amounts to systematic copyright infringement, even under the new and permissive EU law provisions, to which we now turn.

---

[21] See,e.g.,https://www.bakerlaw.com/services/artificial-intelligence-ai/case-tracker-artificial-intelligence-copyrights-and-class-actions/.

[22] Cf. European Parliament resolution of 20 October 2020 on intellectual property rights for the development of artificial intelligence technologies, 2020/2015(INI), par. 15.

[23] Directive 96/9/EC of the European Parliament and of the Council of 11 March 1996 on the legal protection of databases ("Database Directive"), OJ L 77, 27.3.1996, p. 20 – 28.

[24] As clarified by the Court of Justice of the EU in the *Ryanair* case: CJEU, 15 January 2015, case C-30/14 – *Ryanair*, ECLI:EU:C:2015:10.

A potential regulatory solution to ensure the lawful use of training datasets would involve applying the text and data mining (TDM) exception provided by Directive 2019/790/EU (DSMD)[26] to the training of LLMs. Indeed, Article 2(2) DSMD defines text and data mining as "any automated analytical technique aimed at analyzing text and data in digital form to generate information which includes but is not limited to patterns, trends and correlations". Considering that the training of LLMs certainly encompasses (although it likely extends beyond) automated analysis of textual and data content in digital format to generate information, an argument could be made that such activity falls within the definition provided by the DSM Directive (Dermawan, n.d.). However, the application of the TDM exception in the context of LLMs training raises non-trivial issues (Pesch and Böhme 2023) (Hacker 2021) (see also, more generally, Geiger, Frosio, and Bulayenko 2018; Rosati 2018).

Firstly, where the TDM activity is not carried out by research organizations and cultural heritage institutions for scientific research–e.g., by private companies and/or for commercial purposes–it is permitted under Article 4(3) DSMD only on condition that the use of works and other protected materials "has not been expressly reserved by their right-holders in an appropriate manner, such as machine-readable means in the case of content made publicly available online". This condition underscores our earlier note on the need for LLMs to automatically analyze the Terms and Conditions of websites and online databases.

Secondly, a further element of complexity is that Article 4(2) DSMD stipulates that the reproductions and extractions of content made under Article 4(1) may only be retained "for as long as is necessary for the purposes of text and data mining". In this sense, if one interprets the TDM exception to merely cover the training phase of LLMs (as separate from the validation and testing phases), LLMs should delete copyrighted content used during training immediately after its use. Consequently, these materials could not be employed to validate or test LLMs. In this perspective, to make the text and data mining exception more effective in facilitating LLM development, it is advisable to promote a broad normative interpretation of "text and data mining", encompassing not only the training activity in the strict sense but also the validation and testing of the LLM.

Thirdly, the exception covers only reproductions and extractions, but not modifications of the content–which will often be necessary to bring the material into a format suitable for AI training. Finally, according to Article 7(2) DSMD, the three-step test (Geiger, Griffiths, and Hilty 2008) contained in Article 5(5) of the InfoSoc Directive 2001/29/EC restricts the scope of the TDM exception. According to this general limit to copyright exceptions, contained as well in international treaties (Oliver 2001; Griffiths 2009), such exceptions apply only "in certain special cases which do not conflict with a normal exploitation of the work or other subject-matter and do not unreasonably prejudice the legitimate interests of the rightholder." Importantly, this suggests that the TDM exception cannot justify reproductions that lead to applications that substitute, or otherwise significantly economically compete with, the protected material used for AI training. However, this is, arguably, precisely what many applications are doing (Marcus and Southen 2024). It remains unclear, however, to what extent the three-step-test limits individual applications of the TDM exception in concrete cases before the courts, as opposed to being a general constraint on Member States' competence to curtail the ambit of copyright (Griffiths 2009, 3–4).

As mentioned, legal proceedings have recently been brought in the United States and the EU to contest copyright infringement related to materials used in the training phase by AI systems[27]. While the outcomes of such cases are not necessarily predictive of how analogous cases might be resolved in the EU–for example, in the US the fair use doctrine could be invoked (Gillotte 2020), which lacks exact equivalents in the legal systems of continental Europe–it will be intriguing to observe the approach taken by courts across the Atlantic. Note, particularly, that these cases may, among other things, be decided by the extent to which AI systems substitute for, i.e., compete with, the materials they were trained on (so-called transformativeness, see, e.g., (Henderson et al. 2023), a consideration that parallels the debate mentioned above in EU law on the interpretation of the three-step-test (and its transposition into Member State law (Griffiths 2009, 3–4).

*2) Output generation.*
It is now worth focusing on the legal issues raised by the generation of outputs by LLMs. In this regard, two different aspects must be primarily addressed: the legal relationship between these outputs and the materials used during the training of LLMs, and the possibility of granting copyright or patent protection to these outputs.

As for the first aspect, it is necessary to assess whether LLM-generated outputs: (a) give rise to the potential infringement of intellectual property rights in the pre-existing materials, (b) qualify as derivative creations based on the pre-existing materials, or (c) can be regarded as autonomous creations, legally independent from the pre-existing materials.

An answer to this complex legal issue could hardly be provided in general and abstract terms, requiring proceeding with a case-by-case assessment, i.e., by comparing a specific LLM-generated output with one or more specific pre-existing materials. Such a comparison could in principle be conducted by applying the legal doctrines currently adopted by courts in cases of copyright or patent infringement (or, when appropriate, the legal doctrines adopted to assess whether a certain work/invention qualifies as a derivative work/invention). In this perspective, indeed, the circumstance that the output is generated by a human creator or an AI system does not make a significant legal distinction, except in terms of identifying the subject legally accountable for the copyright infringement.

---

[27] See, e.g., Z. Small, "Sarah Silverman Sues OpenAI and Meta Over Copyright Infringement", The New York Times, 10 July 2023, available at: https://www.nytimes.com/2023/07/10/arts/sarah-silverman-lawsuit-openai-meta.html; B. Brittain, "Lawsuit says OpenAI violated US authors' copyrights to train AI chatbot", Reuters, 29 June 2023, available at: https://www.reuters.com/legal/lawsuit-says-openai-violated-us-authors-copyrights-train-ai-chatbot-2023-06-29/.

In general terms, however, the use of protected materials in the training of an LLM does not imply, *per se*, that the LLM-generated outputs infringe upon the intellectual property rights in these materials[28] or qualify as derivative creations thereof. Broadly speaking, an LLM-generated output could infringe upon legal rights in two main ways. First, if the output exhibits substantial and direct similarities to legally protected elements of pre-existing materials, it would likely violate the (reproduction) right of those materials. Second, if the legally protected aspects or elements of the pre-existing materials appear in the LLM output through indirect adaptations or modifications, always unauthorized, then this output would likely qualify as a derivative creation from the pre-existing materials (Gervais 2022; Henderson et al. 2023). For instance, the fact that a text generated by an LLM shares the same style as the works of a specific author (as would occur if a prompt such as "write a novel in the style of Dr. Seuss" were used) would not imply, *per se*, an infringement of the intellectual property rights of that author. This is because, in most European legal systems, the literary or artistic style of an author is not an aspect upon which an exclusive right can be claimed.

If, by contrast, an infringement is found in an LLM output, the person prompting the LLM would first and foremost be liable because she directly brings the reproduction into existence. However, LLM developers might, ultimately, also be liable. The Court of Justice of the European Union (CJEU) has recently determined that if platforms fail to comply with any of three distinct duties of care, they will be directly accountable for violations of the right to publicly communicate a work.[29] These duties amount to i) expeditiously deleting it or blocking access to infringing uploads of which the platform has specific knowledge; ii) putting in place the appropriate technological measures that can be expected from a reasonably diligent operator in its situation to counter credibly and effectively copyright infringements if the platform knows or ought to know, in a general sense, that users of its platform are making protected content available to the public illegally via its platform; iii) not providing tools on its platform specifically intended for the illegal sharing of such content and not knowingly promoting such sharing, including by adopting a financial model that encourages users of its platform illegally to communicate protected content.[30] These duties could–mutatis mutandis–be transposed to LLM developers concerning the right of reproduction (Nordemann 2024, 2023), although such transposition may not be so straightforward. However, this would make good sense, both from a normative perspective encouraging active prevention of copyright infringement and from the perspective of the coherence of EU copyright law across technical facilities.[31]

---

[28] However, some cases might pose more challenges than others: consider, e.g., the case where an AI system is used to create works that involve existing fictional characters (who are *per se* protected).

[29] CJEU, Joined Cases C-682/18 and C-683/18, YouTube vs. Cyando, ECLI:EU:C:2021:503.

[30] CJEU, Joined Cases C-682/18 and C-683/18, YouTube vs. Cyando, ECLI:EU:C:2021:503, para. 102. The latter point addresses specifically piracy platforms, not YouTube (para. 96 and 101).

[31] In his case, one would further have to investigate if Art. 17 DSMD constitutes a lex specialis to the more general *Cyando* case (Geiger and Jütte 2021; Leistner 2020).

A distinct and further legal issue arises when an LLM-generated output can be regarded as an autonomous creation, legally independent from the pre-existing materials. In this scenario, the question pertains to whether such output may be eligible for protection under IP law, specifically through copyright (in the case of literary, artistic, or scientific works) or through patent protection (in the case of an invention) (Engel, 2020; Hristov, 2016; Klawonn, 2023; Varytimidou, 2023).

As mentioned at the beginning of this paragraph, the fundamental legal problem, here, stems from the anthropocentric stance taken by intellectual property law. While international treaties and EU law do not explicitly state that the author or inventor must be human, various normative hints seem to support this conclusion. In the context of copyright, for instance, for a work to be eligible for protection, it must be original, i.e., it must constitute an author's intellectual creation[32]. This requirement is typically interpreted, also by the Court of Justice of the EU, as the work needed to reflect the author's personality (something that AI lacks, at least for now). Patent law takes a less marked anthropocentric approach, but even here, the so-called inventive step–which, together with novelty and industrial applicability, is required for an invention to be patentable–is normatively defined in terms of non-obviousness to a person skilled in the art[33]. The very existence of moral rights (such as the so-called right of paternity) safeguarding the personality of the author or inventor suggests that the subject of protection can only be human.

Considering these succinct considerations, we can return to the initial question, namely whether an LLM-generated output may be eligible for protection under intellectual property law.

The answer to this question is relatively straightforward when the LLM constitutes a mere instrument in the hands of a human creator, or, to put it differently, when the creative outcome is the result of predominantly human intellectual activity, albeit assisted or enhanced by an AI system. In such a scenario, the European Parliament has stressed that where AI is used only as a tool to assist an author in the process of creation, the current IP framework remains fully applicable[34]. Indeed, as far as copyright protection is concerned, the Court of Justice of the EU has made clear in the *Painer* case[35] that it is certainly possible to create copyright-protected works with the aid of a machine or device. A predominant human intellectual activity can be recognized, also based on the CJEU case law, when the human creator using an LLM makes free and creative choices in the phases of conception, execution, and/or redaction of the work (Hugenholtz and Quintais 2021).

A similar conclusion can be drawn regarding the patent protection of inventive outcomes generated with the support of an LLM (Engel 2020). In this perspective, as noted by some scholars, it would likely be necessary to adopt a broader interpretation of the inventive step requirement, which should be understood in terms of non-obviousness to a person skilled in the art assisted by AI, i.e., an AI-aided human expert (Ramalho 2018; Abbott 2018).

---

[32] Cf. Art. 3(1) of the Database Directive; Art. 6 of the Directive 2006/116/EC of the European Parliament and of the Council of 12 December 2006 on the term of protection of copyright and certain related rights ("Term Directive"), OJ L 372, 27.12.2006, p. 12 – 18; Art. 1(3) of the Directive 2009/24/EC of the European Parliament and of the Council of 23 April 2009 on the legal protection of computer programs, OJ L 111, 5.5.2009, p. 16 – 22.

[33] Cf. Art. 56 of the European Patent Convention.

[34] Cf. European Parliament resolution of 20 October 2020 on intellectual property rights for the development of artificial intelligence technologies, 2020/2015(INI), par. 15.

[35] CJEU, 1 December 2011, case C-145/10, *Painer*, ECLI:EU:C:2011:798.

An opposite conclusion is often reached when the LLM operates in a substantially autonomous manner. For the sake of clarity, it is necessary to explain the meaning of "autonomous" as used in this context (Dornis 2021). Obviously, in the current state of technology, some degree of human intervention–at the very least in the form of prompts–will always be necessary for an LLM to generate any output. However, the mere formulation of a prompt by a human being is likely insufficient to recognize a substantial human contribution to the creative output generated by the LLM. The fundamental legal aspect is that a notable human contribution must be discernible not in the broader creative process, but specifically in the resulting creative outcome. This condition is not met when human intervention merely involves providing a prompt to an LLM or even when minor modifications, legally insignificant, are made to the creative outcome generated by the LLM (e.g., minor editing of an LLM-generated text). By contrast, a level of IP protection might be appropriate for significant modifications made to the text produced by the LLM.

The conclusion above, which argues against copyright or patent protection for contents generated by LLMs in a substantially autonomous manner, finds confirmation in the positions taken on this issue by, e.g., the US Copyright Office,[36] affirmed by the United States District Court for the District of Columbia,[37] and the European Patent Office[38]. Furthermore, such a conclusion is consistent with the fundamental rationale of intellectual property of promoting and protecting human creativity, as also reflected at the normative level.[39]

However, some authors have observed (sometimes with critical undertones) that a rationale for protecting LLMs autonomously generated content is the need to protect investments, made by individuals and/or organizations, aimed at bringing creative products to the market (Hilty, Hoffmann, and Scheuerer 2021; Geiger, Frosio, and Bulayenko 2018).

In this case, the further issue of determining to whom such intellectual property rights should be granted emerges. Some national legislations–not coincidentally, following the common law tradition, which exhibits a less pronounced anthropocentric character compared to civil law tradition–acknowledge the possibility of protecting computer-created works (Goold 2021)–i.e. works "generated by computer in circumstances such that there is no human author of the work,"[40]– granting the copyright to the person "by whom the arrangements necessary for the creation of the work are undertaken"[41]. The identity of such a person, however, remains somewhat unclear, as this could be, depending on the circumstances, the developer of the LLM, its trainer, or its user, possibly even jointly (Guadamuz 2021).

---

[36] On 16 March 2023 the US Copyright Office issued formal guidance on the registration of AI-generated works, confirming that "copyright can protect only material that is the product of human creativity": see Federal Register, Vol. 88, No. 51, March 16, 2023, Rules and Regulations, p. 16191.

[37] United States District Court for the District of Columbia [2023]: Thaler v. Perlmutter, No. 22-CV-384-1564-BAH.

[38] On 21 December 2021 the Legal Board of Appeal of the EPO issued a decision in case J 8/20 (DABUS), confirming that under the European Patent Convention (EPC) an inventor designated in a patent application must be a human being.

[39] Cf. recital no. 10 of Directive 2001/29/EC of the European Parliament and of the Council of 22 May 2001 on the harmonisation of certain aspects of copyright and related rights in the information society

[40] Cf. Sec. 178 of the UK Copyright, Designs and Patents Act 1988 ("CDP Act").

In civil law systems, while awaiting a potential *ad hoc* regulatory intervention, a possible solution could involve applying to LLM-generated outputs the same principle that applies to works and inventions created by an employee within the scope of an employment contract. In such cases, in most EU legal systems, copyright or patent rights are vested in the employer. Similarly, in situations where the "employee" is artificial, the intellectual property right could be granted to the user of an LLM during entrepreneurial endeavours (Spedicato 2019).

## 5. Cybersecurity

Cybersecurity is a complex and, in the current geopolitical environment marked by armed and non-armed conflicts in many parts of the world, increasingly urgent matter. The EU has tackled this area with a range of instruments and provisions that apply, to varying degrees, to Generative AI models, too.

*1) The Cyber Resilience Act and the AI Act.*
While the GDPR, in Art. 32, does mandate state-of-the-art cybersecurity measures for any personal data processing, this provision does not, at least not easily, apply to industrial data (Purtova 2018) – which is often the target of cyberattacks, however.

This gap is supposed to be filled by the Cyber-Resilience Act (CRA), recently approved by the EU Parliament. It introduces cybersecurity measures for digital products across Europe. Targeting both hardware and software, the act mandates that Products with Digital Elements (PDEs) adhere to certain cybersecurity standards from design to deployment. A PDE is defined as 'a software or hardware product and its remote data processing solutions, including software or hardware components being placed on the market separately' (Article 3(1) CRA). Hence, AI systems will generally constitute PDEs, to the extent that they are placed on the market in the EU.

The CRA establishes a comprehensive framework to bolster cybersecurity measures across the European Union. It introduces a staggered approach to securing PDEs, starting with Article 6, which mandates that all PDEs must meet basic cybersecurity requirements to enter the EU market. These *essential requirements* are outlined in Annex I of the CRA and adopt a risk-based methodology. They encompass a wide range of measures including conducting cybersecurity risk assessments to eliminate known vulnerabilities, implementing exploration and mitigation systems, ensuring security by default, providing cybersecurity updates automatically, protecting against unauthorized access, ensuring the confidentiality and integrity of data, requiring incident reporting, and maintaining resilience against DDoS attacks. Additionally, it necessitates ongoing responsibilities throughout the product's lifecycle, such as promptly addressing emerging vulnerabilities, conducting regular security testing, and disseminating security patches swiftly.

---

[41] Cf. Sec. 9(3) of the CDP Act. Similarly, Sec. 11 of the 1997 Copyright Ordinance (Cap. 528) of Hong Kong and Art. 2 of the 1994 New Zealand Copyright Act.

For products deemed as 'important PDEs,' Article 7 stipulates that they must adhere to more stringent requirements, including undergoing *conformity assessments*. This classification is determined based on a specified list in Annex III, which includes important components like operating systems, browsers, personal information management systems, cybersecurity-related systems, and password managers. The integration of AI in any of these listed products automatically subjects the AI models to these enhanced cybersecurity protocols, ensuring a robust defense mechanism is in place against potential cyber threats.

'Critical PDEs,' under the scrutiny of Article 8, are required to implement *substantial cybersecurity* measures. The CRA empowers the Commission to designate what constitutes a critical PDE through delegated acts, referencing an exhaustive list of products that are integral to cybersecurity infrastructure, such as hardware devices with security boxes, smart meter gateways, and smart cards. The use of AI within these specified settings mandates compliance with the substantial cybersecurity framework. This constitutes the highest security level; however, Member States may establish even more stringent obligations for products used in national security or defense. The CRA's dynamic structure, allowing for the updating of Annexes by the Commission, ensures that the legislative framework can, at least in theory, adapt to the rapidly evolving cyber threat landscape and technological advancements.

Although broadly encompasses AI systems under the category of PDEs, the CRA specifically delineates targeted requirements for high-risk AI systems in accordance with the classification set forth in the AIA (Article 8 CRA). To obtain a declaration of conformity, such products must comply with the CRA's essential requirements as detailed in Annex I. As mentioned, this encompasses a range of measures; data processing should be limited strictly to what is necessary for the product's intended purpose, emphasizing data minimization.

Hence , the CRA does not explicitly address Generative AI or LLMs. This gap likely stems from the CRA's alignment with an earlier version of the AIA that did not encompass Generative AI or LLMs. Yet, interpreting the CRA legislator's intent as if they wanted to specifically target the most potentially hazardous AI systems through Article 8 and Annex I, and to maintain systemic coherence within the EU legal framework — especially in alignment with the AIA — it becomes evident that the CRA may benefit from adjustments to explicitly encompass Generative AI and align it with the requirements in the AI Act.

Adapting the CRA to explicitly include Generative AI should be relatively straightforward. The AIA has already laid down a risk-tiered classification and specific regulations for Generative AI (i.e., GPAI). This pre-existing framework offers a clear pathway for incorporating Generative AI into the CRA, potentially through the European Commission's delegated acts. Such integration would enhance the CRA's effectiveness in governing AI technologies and align it more closely with the evolving landscape of AI and its potential risks, thereby reinforcing the EU's commitment to a comprehensive and harmonized legal framework for AI regulation.

Importantly, the AIA currently only mandates cybersecurity measures for high-risk systems (Art. 15) and for GPAI with systemic risk (Art. 55). The Joint Research Centre has issued helpful guidance for interpreting and implementing cybersecurity in the context of AI systems (Joint Research Centre (European Commission) et al. 2023). However, in our view, the regulatory framework in the AIA fails to mirror the fundamental importance of cybersecurity in our age. Generative AI models, in particular, are bound to become new building blocks for literally thousands of derived apps and products, functioning much like a new operating system in some respects. Hence, a backdoor created via insufficient cybersecurity will potentially enable attackers to exploit vulnerabilities in a range of derivative products. Therefore, economic efficiency (patching vulnerabilities once upstream instead of manifold times downstream) and prudence argues for stringent and obligatory cybersecurity measures for all GPAI, not only the largest ones ("systemic risk"), such as GPT-4 or Gemini. Strategic rivals, both nation-states and non-state actors, will be actively trying to exploit any vulnerabilities in advanced AI systems, particularly if the systems are widely used and integrated. Not addressing these threats for all GPAI seems naïve at best, and irresponsible in the current and future geopolitical climate.

Hence, in our view, general-purpose AI systems should be included under the categories of Annex III CRA. This would ensure that they fulfill most stringent cybersecurity requirements, including conformity assessments. In the current geopolitical climate, and with the importance of foundation models starting to rival those of operating systems (which are included in Annex III CRA already), this seems like a sensible update. In addition, a link between Article 55 AI Act and the CRA should be included for the cybersecurity requirements concerning systemic risk GPAIS, mirroring the integration of cybersecurity obligations for high-risk AI systems into the AI Act (Article 12 CRA).

In short, generative AI legislation needs a critical cybersecurity patch. Below, we show that several specific cybersecurity concerns remain unaddressed by the current regulatory landscape, including the AIA, CRA, and broader EU legislation.

*2) Adversarial attacks.*
The complexity and high dimensionality of Generative AI models make them particularly susceptible to adversarial attacks, i.e., attempts to deceive the model and induce incorrect outputs – such as misclassification – by feeding carefully crafted, adversarial data. Cybersecurity is a national competence (Cybersecurity Act, Recital 5) but joint efforts to address it should still be pursued at the EU level, going beyond the general principle of AI robustness. Importantly, the AIA mandates high-risk systems to implement technical measures to prevent or control attacks trying to manipulate the training dataset ('data poisoning'), inputs designed to cause the model to make a mistake ('adversarial examples'), or model flaws (Article 15, AIA). The EU's Joint Research Centre has recently unveiled a comprehensive guidance document on cybersecurity measures in the context of AI and LLMs (Joint Research Centre (European Commission) et al. 2023). The European Parliament's draft legislation adds another layer. Article 28b asks GPAI providers to build in "appropriate cybersecurity and safety" safeguards, echoing the two-tiered approach tentatively agreed upon in the trilogue (Hacker 2023c). However, effectively countering adversarial attacks requires careful prioritization and targeting within any AI system, not just high-risk ones.

The AIA's risk levels, based on the likelihood of an AI system compromising fundamental legal values, are not a reliable predictor of vulnerability to adversarial attacks. Some AI deemed as high-risk by the AIA, e.g., for vocational training, may not have those technical traits that trigger adversarial attacks, and vice versa. Therefore, the AIA, and by extension the CRA which relies on its risk classification, should provide, through supplementary implementation acts, technical safeguards that are proportionate to the attack-triggers of a specific LLM, independently of the AIA risk levels. Attack-triggers include model complexity, overfitting, linear behaviour, gradient-based optimization, and exposure to universal adversarial triggers like input-agnostic sequences of tokens (Wallace et al. 2021). Finally, novel methods to counter adversarial attacks might involve limiting LLM access to trusted users or institutions and restricting the quantity or nature of user queries (Goldstein et al. 2023).

*3) Misinformation.*
LLMs can disseminate misinformation, easily, widely, and at a low cost by attributing a high probability to false or misleading claims. This is mainly due to web-scraped training data containing false or non-factual information (e.g., fictional), which lacks truth value when taken out of context. Other times, an opinion reflecting the majority's viewpoint is misrepresented as truth, despite not being verified facts. Misinformation may facilitate fraud, scams, targeted and non-targeted manipulation (e.g., during elections) (AlgorithmWatch AIForensics 2023), and cyber-attacks (Weidinger et al. 2021; Ranade et al. 2021).

A concerning aspect of natural language processing (NLP) in general is the phenomenon of "hallucinations". It refers to the generation of seemingly plausible text that diverges from the input data or contradicts factual information (Ye et al. 2023). These hallucinations arise due to the models' tendency to extrapolate beyond their training data and synthesize information that aligns with their internal patterns, even if it is not supported by evidence. As a result, while NLP models may produce texts that demonstrate coherence, linguistic fluidity, and a semblance of authenticity, their outputs often lack fidelity to the original input and/or are misaligned with empirical truth and verifiable facts (Ji et al. 2023). This can lead to a situation where uncritical reliance on LLMs results in erroneous decisions and a cascade of negative consequences (Zhang et al. 2023), including the spread of misinformation, especially if false outputs are shared without critical evaluation.

There are different kinds of LLMs' hallucinations (Ye et al. 2023) but we cannot discuss them here in detail. In the recent generation of LLMs–e.g., GPT4 and Bard–the 'Question and Answer' kind is particularly frequent. These hallucinations manifest due to the models' tendency to provide answers even when presented with incomplete or irrelevant information (Ye et al. 2023; Adlakha et al. 2023). A recent study found that hallucinations are particularly common when using LLMs on a wide range of legal tasks (Dahl et al. 2024).

EU legislation lacks specific regulations for misinformation created by Generative AI. As LLMs become increasingly integrated into online platforms, expanding the Digital Services Act (DSA) to include them, and mandating online platforms to prevent misinformation, seems the most feasible approach. Also, the project to strengthen the EU Code of Practice on Disinformation (2022) can contribute, though its voluntary adherence reduces its overall effectiveness. Tackling LLM-generated misinformation requires updating both the AIA and the DSA. The DSA contains a range of provisions that can be fruitfully applied to LLMs: e.g., Article 22, which introduces "trusted flaggers" to report illegal content to providers and document their notification (Hacker, Engel and Mauer 2023).

However, it is essential to broaden the DSA's scope and the content subject to platform removal duty, which currently covers only illegal content, as LLM-generated misinformation may be completely lawful (Berz, Engel, and Hacker 2023). Being the most technology-focused regulation, the AIA, or its implementing acts, should tackle design and development guidelines to prevent LLMs from spreading misinformation. Normative adjustments should not focus only on the limitation of dataset size but also explore innovative strategies that accommodate LLMs' data hunger. Some measures might be the same (or similar to those) mentioned for adversarial attacks — restricting LLM usage to trusted users with limited interactions to prevent online misinformation proliferation[42] — while others may include innovative ideas like fingerprinting LLM-generated texts, training models on traceable radioactive data, or enhancing fact sensitivity using reinforcement learning techniques (Goldstein et al. 2023).

Specific solutions to address hallucinations in LLMs are crucial for mitigating the spread of misinformation and should be employed in policy-related applications. Numerous approaches have been proposed in the literature to address this challenge (Tonmoy et al. 2024; Ye et al. 2023). Some of these solutions are broad strategies that focus on optimizing dataset construction, such as implementing a self-curation phase within the instruction construction process. During this phase, the LLM identifies and selects high-quality demonstration examples (candidate pairs of prompts and responses) to fine-tune the underlying model to better follow instructions (Li et al. 2023). Other strategies address the alignment of LLMs with specific downstream applications–which can benefit from supervised fine-tuning (Chung et al. 2022)–as hallucinations often arise from discrepancies between the model's capabilities and the application's requirements (Ye et al. 2023).

Other approaches are narrower and focused on specific techniques, such as prompt engineering, to optimize the output generated by LLMs. This includes incorporating external authoritative knowledge bases (retrieval-augmented generation) (Kang, Ni, and Yao 2023) or introducing innovative coding strategies or faithfulness-based loss functions (Tonmoy et al. 2024).[43]

Another technical solution to mitigate hallucinations in LLMs worth considering is the Multiagent Debate approach, where multiple LLMs engage in an iterative process of proposing, debating, and refining their responses to a given query (Du et al. 2023). The aim is to achieve a consensus answer that is not only more accurate and factually correct but also preserves the richness of multiple perspectives (Ye et al. 2023). This approach draws inspiration from judicial techniques, particularly cross-examination, to foster a more rigorous examination of the LLMs' responses (Cohen et al. 2023).

*4) Ways forward: NIS2.*
The provisional agreement on the EU's updated Network and Information Systems Directive (NIS2 Directive) signifies a major update to the bloc's cybersecurity framework, set to supersede the initial Network and Information Systems Directive.

---

[42] For instance, the draft legislative proposal of the European Parliament requires that the provider of a foundation model (now GPai shall demonstrate the reduction and mitigation of reasonably foreseeable risks to democracy and the rule of law (Article 28b).

[43] Which basically means establishing a metric to measure faithfulness, that is, the extent to which a model's outputs align with the input data or established truths.

With its formal adoption expected soon, NIS2 extends coverage to more sectors and entities (Annexes I and II).

NIS2 mandates that designated essential and important entities adopt measures across technical, operational, and organizational domains to address risks to their network and information systems (Article 3 NIS2). These precautions aim to either prevent or mitigate the effects of cyber incidents on users, maintaining security proportionate to assessed risks (Article 21 (1) NIS2). It also introduces requirements for enhancing supply chain security, focusing on the relationship with direct suppliers and service providers, to shield against cyber incidents.

The NIS2 Directive significantly expands cybersecurity measures beyond those of its predecessor, the NIS Directive, covering additional sectors and entities. This makes it highly relevant for those in Generative AI, including the digital infrastructure and services sectors, which would naturally involve companies working with (Generative) AI. Additionally, NIS2 mandates quick incident reporting, requiring entities to inform authorities within 24 hours of certain cybersecurity incidents (Article 23 point 4(a) NIS2). This is crucial for the AI sector, where only a rapid response to security breaches can mitigate the consequences, such as exploiting AI vulnerabilities or malicious AI activities.

In this context, the interplay between the NIS2 Directive and the CRA is crucial, particularly in how NIS2 can enhance or compensate for the CRA's limitations. For instance, the CRA proposal focuses on ensuring high cybersecurity standards for *products* (with digital elements, i.e., PDEs), yet it does not fully extend these standards to *services*, except for "remote data processing solutions" (Article 3 CRA) (Eckhardt and Kotovskaia 2023). This gap could leave various generative AI models without adequate cybersecurity coverage, especially when these technologies are integrated into products or services beyond remote data processing. This includes scenarios where Generative AI and LLMs are part of more complex systems or services that offer decision-making, content generation, or predictive analytics. The NIS2 Directive takes a broader approach by targeting essential and significant entities, including cloud computing service providers. This implies that if generative AI and LLMs are offered through cloud services that meet the criteria for being considered essential or significant (e.g., due to their size or the critical nature of the services they provide), they would fall under the cybersecurity and incident notification requirements of NIS2.

## 6. Conclusion

State-of-the-art Generative AI models in general, and LLMs in particular, exhibit high performance across a broad spectrum of tasks, but their unpredictable outputs raise concerns about the lawfulness and accuracy of the generated content. Overall, EU does not seem adequately prepared to cope with these novelties. Policy proposals include updating current and forthcoming regulations, especially those encompassing AI more broadly, as well as the enactment of specific regulations for Generative AI. In this article, we have offered an overall analysis of some of the most pressing challenges and some suggestions about how to address them. The broader point about how best to proceed in the development of a very complex and yet entirely coherent EU architecture of "digital laws" remains to be addressed. Ultimately, technological solutions may help if we start asking not only what the law can do for the development of socially preferable AI, but also what AI can do to improve the relevance, coherence and timeliness of the law. But this is a topic beyond the scope of this article.

---

[i]Authors have worked on different sections according to the following division: Claudio Novelli has worked on Sections 1, 2, 3, 5, 6; Federico Casolari has worked on Section 2 ; Philipp Hacker has worked on Sections 2, 3, 4, 5; Giorgio Spedicato has worked on Section 4; Luciano Floidi has worked on Sections 1, 3, 6.